\documentclass[a4paper,11pt]{article}
\pdfoutput=1 
\usepackage{amsmath}
\usepackage{mathtools}
\usepackage{tabu}
\usepackage{aas_macros}
\usepackage{jcappub} 
\usepackage[T1]{fontenc} 
\usepackage{graphicx,eurosym,wrapfig,layout,paralist,color,gensymb}
\usepackage[normalem]{ulem}

\newcommand{\kms}{km~s$^{-1}$}
\newcommand{\bctr}{\begin{centering}}
\newcommand{\ectr}{\end{centering}}
\newcommand{\beq}{\begin{equation}}
\newcommand{\eeq}{\end{equation}}
\newcommand{\bea}{\begin{eqnarray}}
\newcommand{\eea}{\end{eqnarray}}
\newcommand{\om}{\Omega_{\rm m}}

\newcommand{\oll}{\Omega_{\rm \Lambda}}

\newcommand{\hinvMpc}{$h^{-1}$Mpc}
\newcommand{\lcdm}{$\Lambda$CDM}
\newcommand{\lsim}{\mbox{$\:\stackrel{<}{_{\sim}}\:$} }

\newcommand{\zbar}{\ensuremath{\bar{z}}}
\title{The need for accurate redshifts in supernova cosmology}

\author[a,b]{\note{Corresponding author.}Josh Calcino$^1$}
\author[a,b]{and Tamara Davis}

\affiliation[a]{The School of Mathematics and Physics, \\  University of Queensland, QLD 4072, Australia}
\affiliation[b]{ARC Centre of Excellence for All-sky Astrophysics (CAASTRO)}

\emailAdd{j.calcino@uq.edu.au}

\abstract{Recent papers have shown that a small systematic redshift shift ($\Delta z\sim 10^{-5}$) in measurements of type Ia supernovae can cause a significant bias ($\sim$1\%) in the recovery of cosmological parameters.  Such a redshift shift could be caused, for example, by a gravitational redshift due to the density of our local environment.
The sensitivity of supernova data to redshift shifts means supernovae make excellent probes of inhomogeneities.  We therefore invert the analysis, and try to diagnose the nature of our local gravitational environment by fitting for $\Delta z$ as an extra free parameter alongside the usual cosmological parameters. 

Using the Joint Light-curve SN Ia dataset we find the best fit includes a systematic redshift shift of $\Delta z = (2.6^{+2.7}_{-2.8}) \times 10^{-4}$.  
This is a larger shift than would be expected due to gravitational redshifts in a standard $\Lambda$-Cold Dark Matter universe (though still consistent with zero), and would correspond to a monopole Doppler shift of about 100 km\,s$^{-1}$ moving away from the Milky-Way.  However, since most supernova measurements are made to a redshift precision of no better than $10^{-3}$, it is possible that a systematic error smaller than the statistical error remains in the data and is responsible for the shift; or that it is an insignificant statistical fluctuation.

We find that when $\Delta z$ is included as a free parameter while fitting to the JLA SN Ia data, the constraints on the matter density shifts to $\Omega_m = 0.313^{+0.042}_{-0.040}$, bringing it into better agreement with the CMB cosmological parameter constraints from Planck.
A positive $\Delta z\sim 2.6\times10^{-4}$ would also cause us to overestimate the supernova measurement of Hubble's constant by $\Delta H_0 \sim $1 kms$^{-1}$Mpc$^{-1}$.  However this overestimation should diminish as one increases the low-redshift cutoff, and this is not seen in the most recent data.
}

\begin{document}
\maketitle

\section{Introduction}
\label{sec:intro}

The cosmological principle has so far been a successful basis for our understanding of the universe. However as technology and observational techniques improve, we must begin to re-evaluate the validity of past assumptions. Although the universe is statistically homogeneous over scales larger than about 100 Mpc \cite{scrimgeour12}, in detail there are large local variations. Matter tends to clump together with intersecting filaments, while leaving large volumes, known as voids, with relatively little matter. This has an impact on our observations, which are inherently location dependent. Gravitational lensing and peculiar velocities are two well studied examples that are generally accounted for when fitting for cosmology. Gravitational redshifts, on the other hand, have generally been overlooked as they were considered negligible.  

However, recent papers have shown that even small gravitational redshifts can have quite a large impact on supernova cosmology \citep{lavallaz2011,marra2013,yu2013,valkenburg2013,wojtak2015}.  A systematic redshift shift of only $\sim 2\times 10^{-5}$ can shift the best fit cosmological parameters by a few percent in a cosmological constant plus cold-dark matter cosmological model ($\Lambda$CDM) \citep{wojtak2015}.  The effect is larger for more complex models, as a dark energy equation of state that differs from $w=-1$ or a dark energy that varies with time, can mimic some of the behaviour of a redshift shift. Futhermore, gravitational redshifts have the potential to delude observers into choosing the incorrect cosmological model for the universe. For example, \cite{valkenburg2013} show that there is an intrinsic limit to how well we can measure cosmological parameters in an inhomogeneous universe and that observers will detect a redshift-dependent $w$ in a flat $\Lambda$CDM universe if they assume the metric surrounding them is FLRW, that is, they assume homogeneity and isotropy. It is important to clarify that though the papers cited above place an emphasis on gravitational redshifts, any systematic redshift bias would be able to produce a similarly significant result.

Gravitational redshifts on the order of $4\times 10^{-5}$ have been measured in galaxy clusters \citep{wojtak2011, dele2012, sadeh2015}, meaning that it is possible that gravitational redshifts are hidden in the supernovae dataset. If this is the case, then we expect that our best fit cosmological parameters derived from supernovae would also be affected. Thus with modern data we need to account for possible gravitational redshifts and evaluate any possible biasing effects, as they may no longer be considered negligible.

Recent cosmological results obtained from the three main cosmological probes have revealed a slight tension in the value of Hubble's constant. Results from Type Ia supernovae disagree with the value of Hubble's constant from the Cosmic Microwave Background (CMB) and Baryon Acoustic Oscillations (BAO) \cite{boss2016}. The value of the Hubble constant determined from the distance ladder (megamasers, Cepheids, and Type 1a supernovae) of $H_0 =73.03 \pm 1.79$ km\,s$^{-1}$Mpc$^{-1}$ \citep{riess2016} disagrees with the Planck 2015 result of $H_0 = 67.51 \pm 0.64$ km\,s$^{-1}$Mpc$^{-1}$ \cite{planck2015}. By introducing a redshift bias parameter we alleviate some of this tension, as we show in Sect~\ref{sec:deponhub}.

The gravitational redshift from individual supernovae in different locations will average out somewhat, because of the diversity of host environments.  However, the one gravitational redshift that would be common to all supernovae we observe is the gravitational redshift due to our own local environment.   
Given the high sensitivity supernovae exhibit as a response from gravitational redshifts, we have the opportunity to probe our local environment in a new way.  In this paper we assess whether the current supernova data is strong enough to reveal whether we live in a local over- or under-density.  We do this by adding a systematic redshift shift of $\Delta z$ as an extra parameter to our cosmological fit. 

\section{Theory}
The Friedmann equations of cosmology are derived using the cosmological principle in the form of a perfect fluid, which is homogeneous and isotropic. This is substituted into the stress-energy-momentum tensor in the Einstein field equations for General Relativity, which then leads to the familiar equations for cosmological models such as $\Lambda$CDM. By definition every location in this model has the same gravitational potential. This assumption remained valid for many purposes until recently, but now more of our observational capabilities are becoming precise enough to be sensitive to the fluctuations about this mean. 

The structure of the universe is defined by the cosmic web of interweaving filaments and large, low density voids. These contrasting environments give rise to a diverse range of gravitational environments. Redshift offsets will be seen whenever the gravitational potential of the emitter and observer differ. The gravitational potentials of most importance are the large super-cluster-sized density fluctuations upon which galaxies are small perturbations, so even galaxies that appear very similar in mass and size can have different gravitational potentials \citep{wojtak2015}. The light emanating from galaxies in large over-densities will become redshifted as it climbs out of the gravitational potential well in which the galaxy resides. If an observer resides in a region of the cosmic web which is under-dense compared to the universal mean, they will observe a systematic redshift bias as the light from distant sources loses energy to reach the observer.

The present day comoving distance of an object at some redshift, $z$, is given by the expression
\begin{equation}
D(\bar{z}) = R_0 \chi(\bar{z}) = c\int^{\bar{z}}_0 \frac{dz}{H(z)},
\label{eq:comoving}
\end{equation}
where $H(z)$ is the Hubble constant as a function of redshift, $c$ is the speed of light, $\chi$ is the comoving coordinate, and $R_0$ is the scale factor at the present day. The redshift $\bar{z}$ in this equation is assumed to be purely cosmological, arising solely from the expansion of the universe. In reality, the redshift that the observer actually measures is a combination of the true cosmological redshift as well as any redshift due to the peculiar velocity of the galaxy ($z_{\rm pec}^{\rm gal}$), the peculiar velocity of the observer ($z_{\rm pec}$), the gravitational redshifts of the galaxy ($z_{\phi}^{\rm gal}$) and observer ($z_{\phi}$), and of course any observational error.  For our purposes we are only interested in a systematic observational error ($z_{\rm sys}$) because random errors are already taken into account in the cosmological fitting. Thus the total redshift the light experiences between source and earth  (i.e.\ geocentric) is,
\begin{equation}
1+z_{\rm geo} = (1+\bar{z})(1+z_{\rm pec})(1+z_{\rm pec}^{\rm gal})(1+z_{\phi})(1+z_{\phi}^{\rm gal}),
\label{eq:zobs}
\end{equation}
and once the systematic observational error has been included the observed redshift would be (without statistical error),
\begin{equation}
1+z_{\text{tot}} = (1+z_{\rm geo})(1+z_{\rm sys}).
\label{eq:ztot}
\end{equation}

Our peculiar motion is easily accounted for by using the CMB as a reference frame. Typically the measured redshift is initially reported in the heliocentric frame, $z_{\rm hel}\approx z_{\rm geo}$, where corrections on the order of $<$30\kms\ have usually been automatically applied to correct for the Earth's motion around the Sun. These are then further transformed using the well-known CMB dipole velocity into the CMB reference frame, where the aim is for the resulting $z_{\text{cmb}}$ to be exactly equivalent to the cosmological redshift $\bar{z}$.   We summarise all these different redshift contributions in Table~\ref{tb:zs}. 

When determining the luminosity distance one subtlety is often overlooked, namely that two different redshifts should be used in different parts of the equation. Luminosity distance is a combination of the true comoving distance that would have been measured in the absence of peculiar velocities and gravitational redshifts, with a factor of $(1+z)$ that occurs due to relativistic beaming and time-dilation (each contributing a factor of $\sqrt{1+z}$).  These latter two effects care about the total relative motion of the source and observer, and therefore should use the geocentric redshift,  
\begin{equation}
D_L = (1+z_{\rm geo}) D(\bar{z}).
\label{eq:DLcorrect}
\end{equation}
This equation is then used for fitting SN Ia and determining the cosmological parameters.  Mistakenly using $\bar{z}$ in both causes only a very small error, but using $z_{\rm geo}$ for both would cause an error that significantly impacts the cosmological parameter estimate \citep{calcino2015}.  

Therefore it is important to convert from the observed redshift to the cosmological redshift ($\bar{z}$) before calculating the comoving distance to the supernova.  Unfortunately we only know one of the terms on the right of Eq.~\ref{eq:zobs}, namely $z_{\rm pec}$ (our velocity with respect to the CMB).  If the other terms are non-negligible then it can bias our cosmological results. 

\section{Methodology}
We use the Joint Light-curve Analysis (JLA) sample \cite{betoule2014} for the basis of this investigation. The dataset contains 740 spectroscopically confirmed SN Ia in a redshift range of $0.01 < z < 1.3$ from the Sloan Digital Sky Survey II (SDSS-II) \cite{sako2014}, the Supernova Legacy Survey (SNLS) \cite{astier2006,sullivan2011}, the Hubble Space Telescope (HST) \cite{riess2007, suzuki2012}, and low-redshift samples \cite{hicken2009,contreras2010,folatelli2010,stritzinger2011,ganeshalingam2013,aldering2002}.

\subsection{The Distance Modulus and Covariance Matrix}
To calibrate SN Ia as standard candles one needs to account for the correlations between the peak luminosity and the width of the SN Ia light curve \citep{phillips1993,phillips1999}, and between dust extinction and colour \citep{tripp1998}.  Several methods have been used to perform this calibration; the JLA sample calculates the distance modulus as,

\begin{equation}\label{eq:muo}
\mu = m_B - (M_B - \alpha X_1 + \beta C)
\end{equation}
where each supernova is described by the combination of $(m_B, X_1, C)$, with $m_B$ representing its apparent rest-frame B-band magnitude, $X_1$ its stretch (light curve witdth), and $C$ its colour, which encapsulates both intrinsic colour variation and the extinction due to the intervening gas and dust.  The parameters $(M_B, \alpha, \beta)$ are common to the entire sample, and are treated as nuisance parameters during the cosmological fit.  It has been found that both $M_B$ and $\beta$ are dependent on the host galaxy properties, and this effect is reconciled by introducing another nuisance parameter $\Delta _M$ such that,
\begin{align}\label{eq:MB}
M_B =
\begin{dcases}
    M_B^1 \qquad &\text{if  } M_{\text{stellar}} < 10^{10}M_{\odot},\\
    M_B^1 + \Delta _M \quad &\text{otherwise}.\\
\end{dcases}
\end{align}
This theoretical distance modulus is, 
\begin{equation}\label{eq:theorymu}
\mu = 5 \log_{10}(d_L)+ 2 5+ 5\log_{10} \left( \frac{c}{H_0} \right),
\end{equation}
where $d_L=D_L H_0/c$ is a function of the cosmological parameters of interest.  A fiducial value of $H_0 = 70$kms$^{-1}$Mpc$^{-1}$ is used, as in \cite{betoule2014}, but since we marginalise over absolute magnitude this fiducial value has no impact on the resulting cosmological parameters. 

There are correlations between the distance moduli of supernovae as a function of their stretch and colour properties, so it is crucial to take these into account during the likelihood analysis. In the JLA analysis the statistical and systematic uncertainties are combined into a single $3N_{\text{SN}} \times  3N_{\text{SN}}$ covariance matrix, \textbf{C}, where $N_{\rm SN}$ represents the number of supernovae in the sample.

This covariance matrix takes into account uncertainties and correlations due to light-curve fitting, calibration, the light-curve model, bias correction, host mass, dust, and peculiar velocities.  These are added to diagonal uncertainties contributed by the dispersion in redshifts, lensing, and coherent scatter.  
The best fit model is then the one that minimises,
\begin{equation}
\chi ^2_{\text{SN}} = \sum\left(\boldsymbol{\hat{\mu}} - \boldsymbol{\mu}_{\text{$\Lambda$CDM}}\right)^T \textbf{C}^{-1}\left(\boldsymbol{\hat{\mu}} - \boldsymbol{\mu}_{\text{$\Lambda$CDM}}\right),
\end{equation}
where $\boldsymbol{\hat{\mu}}$ and $\boldsymbol{\mu}_{\text{$\Lambda$CDM}}$ are the vectorised luminosity distances defined by equations \eqref{eq:muo} and \eqref{eq:theorymu} respectively.

For more detail see Section 5 of \cite{betoule2014}.

In our analysis we choose the simplest viable cosmological model -- a flat $\Lambda$CDM universe --  and add to this an extra parameter which encapsulates a redshift bias.  The flat $\Lambda$CDM model is described by a single parameter, the matter density $\om$ (as well as the Hubble parameter $H_0$, which we effectively marginalise over in supernova analyses). 
 We implement the additional redshift bias parameter, $\Delta z$, by assuming the redshift we {\em actually} observe is 
\begin{equation}
1+z_{\rm obs} =  (1 + z_{\rm geo})(1 + \Delta z).
\end{equation}
And consequently, after correcting for the observed dipole in the CMB we would derive our redshift with respect to the CMB frame to be
\begin{equation}
1 + z_{\rm cmb} = (1+\bar{z})(1+\Delta z).
\end{equation}

Here we are emphasising that our observed redshift, and thus the CMB redshift that is derived from it, deviate slightly from the redshifts that should be used when calculating the luminosity distance specified in equation \eqref{eq:DLcorrect}.   Namely,mwe measure $D_L=(1+z_{\rm obs}) D(z_{\rm cmb})$ but usually analyse it as though we had observed $D_L=(1+z_{\rm geo}) D(\bar{z})$. Here we assess the impact of that assumption on the resulting cosmological parameters and search for a redshift bias by allowing $\Delta z$ to be a free parameter. The full set of parameters that we fit for is therefore $\{\om, \Delta z;\alpha,\beta,M^1_B,\Delta_M\}$, where we marginalise over the parameters after the semicolon.
 
\begin{table}
\caption{Redshift definitions}
\centering
\label{tb:zs}
\resizebox{\textwidth}{!}{%
{\tabulinesep=0.7mm
\begin{tabu}{lll}
\hline \\
$z_{\rm geo}$ & Geocentric & Redshift in Earth's reference frame; differs from $z_{\rm obs}$ by systematic or observational errors.\\
$z_{\rm hel}$ & Heliocentric & Redshift in the Sun's reference frame; correction of $\lsim 30$\kms\ ($z\lsim 0.0001$) from $z_{\rm geo}$. \\
$z_{\rm cmb}$ & CMB &  Redshift in CMB reference frame; correction of $\lsim$360\kms\ ($z\lsim 0.001$) from $z_{\rm hel}$. \\
$z_{\rm obs}$ & Observed & Redshift observed; should be equivalent to $z_{\rm geo}$ in absence of unaccounted for systematics.  \\
$\bar{z}$ & Cosmological & Redshift due to the expansion of the universe; equivalent to $z_{\rm cmb}$ in absence of systematics. \\
$\Delta z$ & Systematic & Systematic redshift offset, due to gravitational redshift or measurement systematic error. \\
$z_{\rm pec}$ & Our peculiar velocity & Redshift due to our peculiar velocity with respect to the CMB.\\
$z_{\rm pec}^{\rm gal}$ & Source peculiar velocity & Redshift due to source's velocity with respect to the CMB. \\
$z_{\rm \phi}$ & Our gravitational & Redshift due to our gravitational potential  \\
$z_{\rm \phi}^{\rm gal}$ & Source gravitational & Redshift due to source's gravitational potential. \\
\hline
\end{tabu}}
}
\end{table}

\section{Results and Discussion}\label{sect:results}

\begin{table}[t]
\centering
\caption{The tabulated results from the analysis on the JLA sample by \cite{betoule2014}, followed by the analysis that we conducted with an additional $\Delta z$ parameter. The slight difference between the original JLA results  \cite{betoule2014} (first row) and our analysis of the same data (second row) arises because we corrected the redshift terms in the luminosity distance (Eq.~\ref{eq:DLcorrect}). Adding $\Delta z$ has a very small effect on the nuisance parameters, but a more significant effect on the parameter of most interest, $\om$. Constraints involving CMB data are obtained by simply using a Gaussian matter density prior  of $\om=0.3121\pm 0.0087$ to reflect the results of Planck \cite{planck2015}.}
\label{tb:res}
\resizebox{\textwidth}{!}{%
{\tabulinesep=0.7mm
\begin{tabu}{lccccccc}
\hline
\hline
 						& $\Omega_m$ 		& $\Delta z\; (\times 10^{-4})$ & $\alpha$ 				& $\beta$ 			& $M^1_B$ 				& $\Delta_M$			&$\chi ^2$/d.o.f.\\ \hline

JLA	\citep{betoule2014}	& $0.295 \pm 0.034$			&  	---				& $0.141 \pm 0.006$	 & $3.101\pm 0.075$			& $-19.05 \pm 0.02$	& 	-0.070 $\pm$ 0.023	&		682.9/735		\\ \hline
JLA					& $0.296 ^{+0.037}_{-0.035}$	&  	---				& $0.141 \pm 0.007$ & $3.107^{+0.086}_{-0.085}$	& $-19.04 \pm 0.03$ & 	-0.070 $\pm$ 0.024	&		682.7/735		\\
JLA $\Delta z$ 			& $0.313^{+0.042}_{-0.040}$ 	& $2.6^{+2.7}_{-2.8}$ 	& $0.142 \pm 0.007$ & $3.111^{+0.086}_{-0.085}$	& $-19.04 \pm 0.03$ &  	-0.070 $\pm$ 0.024	&		681.8/734		\\ 
JLA with CMB &  $0.311^{+0.009}_{-0.009}$ & ---	& $0.141 \pm 0.007$ & $3.101^{+0.086}_{-0.084}$	& $-19.04 \pm 0.02$ &  	-0.071 $\pm$ 0.024	&		683.0/736		\\ 
JLA $\Delta z$ with CMB &  $0.312^{+0.009}_{-0.009}$ & $2.6 \pm 2.5$	& $0.142 \pm 0.007$ & $3.101^{+0.085}_{-0.084}$	& $-19.04 \pm 0.02$ &  	-0.070 $\pm$ 0.024	&		681.8/735		\\ \hline
\end{tabu}}
}
\end{table}

\begin{figure}
	\begin{center}
	\includegraphics[width=0.7\textwidth]{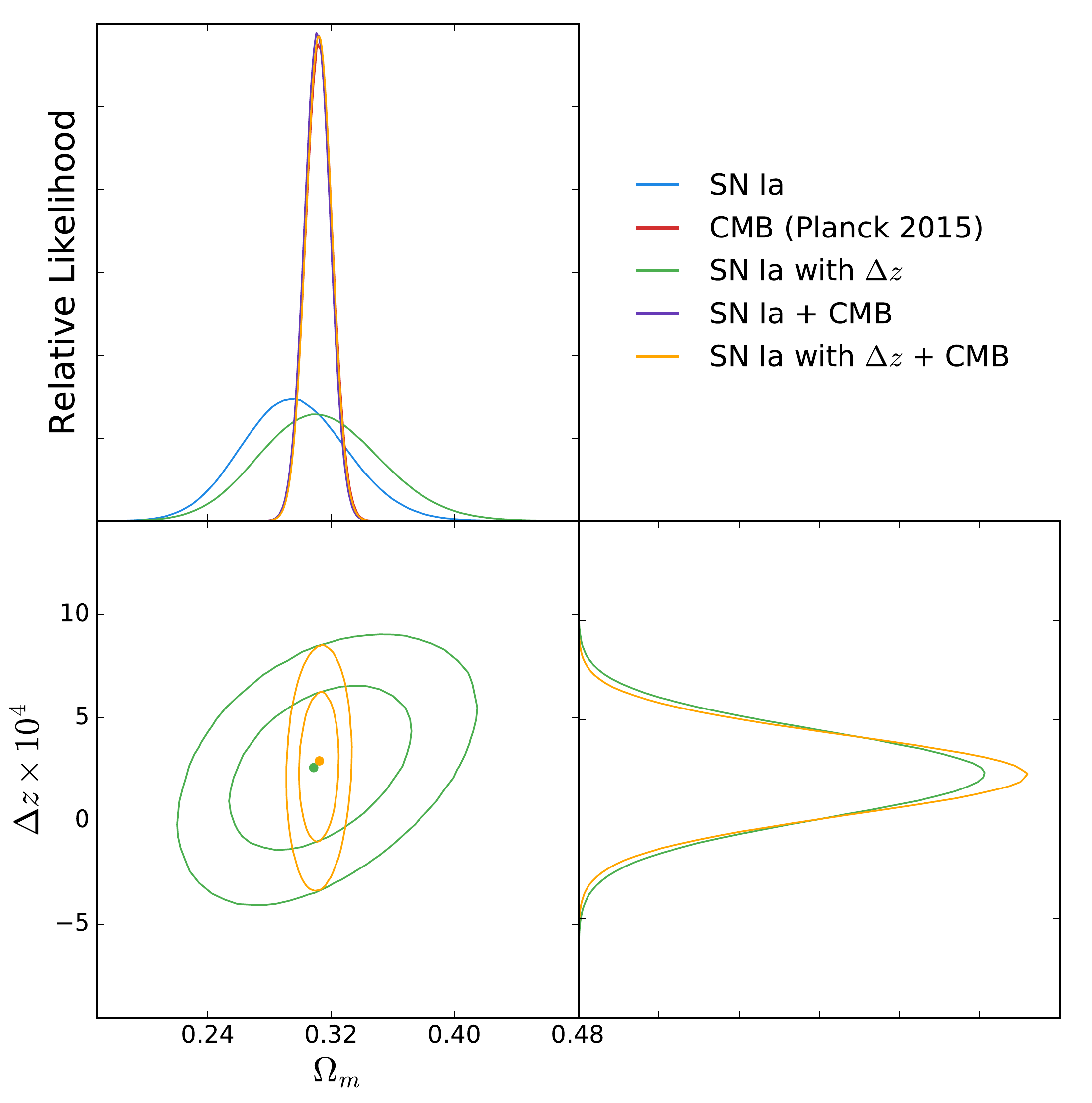}
	\end{center}
	\caption{Results of fitting the $\Lambda$CDM model to supernova data with (green) and without (blue) adding the $\Delta z$ parameter, compared to the matter density measured from the CMB by Planck \citep{planck2015} (red).  Allowing for a non-zero $\Delta z$ naturally moves the best fit from supernovae more in line with the CMB data. The matter density constraints for the CMB, SN Ia + CMB, and SN Ia with $\Delta z$ + CMB overlap each other. Note that no CMB information is used in generating the  SN Ia with $\Delta z$ fit.}
	\label{fig:results}
\end{figure}

The main results of our analysis are shown in Fig.~\ref{fig:results} and Table~\ref{tb:res}.\footnote{
Our results for the basic \lcdm\ fit differ slightly from the those obtained by \cite{betoule2014} when using the same dataset. This difference arises because \cite{betoule2014} do not take into account our Earth's motion when calculating the luminosity distance and use $D_L = (1+\bar{z})D(\bar{z})$.  
Although this does not cause a significant deviation for this data set, we use the ``observed'' (heliocentric) redshift in the first term (assuming the effect of the conversion between geocentric and heliocentric to be negligible). }
The addition of $\Delta z$ to the fit causes a  change in the value of matter density from $\om = 0.296 ^{+0.037}_{-0.035}$ to $\om = 0.313^{+0.042}_{-0.040}$.  

\begin{figure}[t]
	\begin{center}
	\includegraphics[width=1\textwidth]{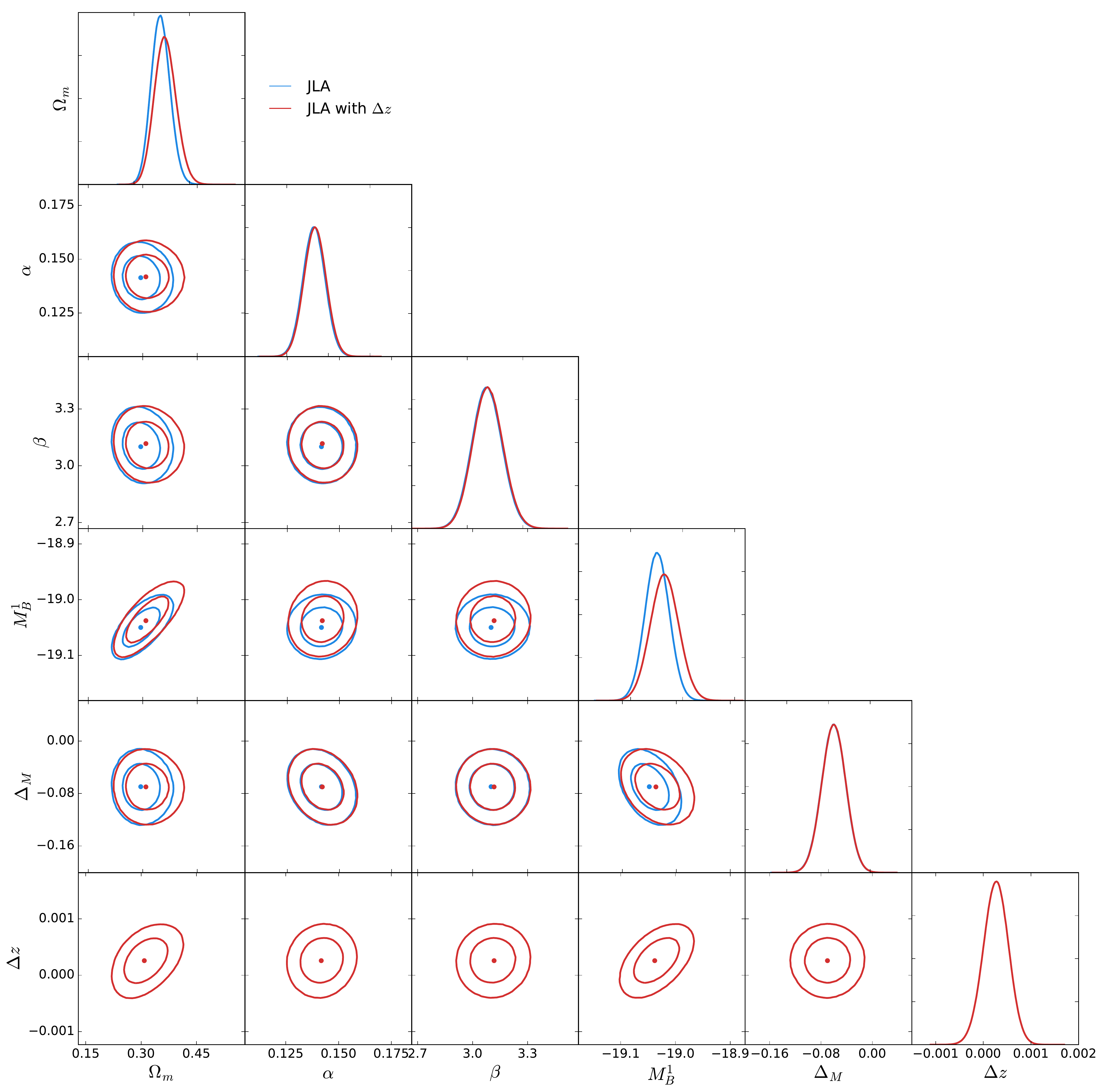}
	\end{center}
	\caption{Full set of parameter contours from fitting the JLA supernova dataset with and without the redshift bias parameter $\Delta z$. The shift in matter density $\om$ is the most significant parameter change. Most of the nuisance parameters are not affected by adding a $\Delta z$.} 
	\label{fig:combinedres}
\end{figure}

We find a best fit value of $\Delta z = (2.6^{+2.7}_{-2.8}) \times 10^{-4}$, which is still consistent with zero to within $1\sigma$. Most of the nuisance parameters are unaffected, but there is a small shift in $M_B^1$ (see Fig.~\ref{fig:combinedres}).  Interestingly, including $\Delta z$ shifts the best fit matter density toward the Planck CMB \citep{planck2015} result of $\Omega_m = 0.3121 \pm 0.0087$. 

Much of the constraining power on $\Delta z$ arises from low-redshift supernovae. Introducing a low-redshift cutoff of $z=0.02$ halves the precision with which we can measure $\Delta z$.

To assess whether complicating our cosmological model with the addition $\Delta z$ is justified we turn to Information Criteria (IC) tests. 
The Bayesian Information Criterion (BIC) \cite{schwarz1978} and the Akaike Information Criterion (AIC) \cite{akaike1974} are two heavily used statistical tests for model selection, and are given by
\begin{eqnarray}
\rm BIC &=& \chi ^2 + k \ln N, \\
\rm AIC &=& \chi ^2 + 2k,
\end{eqnarray}
where $k$ is the number of parameters used in the fitting and $N$ is the number of data points in the sample. For both BIC and AIC the model with the lowest information criterion score is considered to be the better model for explaining a given dataset.  They thus attempt to encapsulate Occam's razor by rewarding models with low $\chi^2$ while penalising models that require many parameters.  An improvement in AIC or BIC of 2 considered positive evidence for the model with lower IC while a difference of 6 is considered strong evidence.  

To test whether the addition of the extra parameter $\Delta z$ is warranted we measure the AIC and BIC with and without $\Delta z$ for both the JLA sample alone and the JLA and CMB data combined. 
The results are presented in Table~\ref{tb:aicbic}. Adding $\Delta z$ increases the IC in both cases, so we find no evidence to support the addition of $\Delta z$ as an extension of the flat $\Lambda$CDM model. This does not necessarily mean $\Delta z\equiv 0$, it simply means the current level of precision of the data is insufficient to show what (if any) $\Delta z$ is required. As the number of SN Ia increase, so too will the importance of any systematic redshift offset introduced by gravitational redshifts \cite{yu2013}. So we should revisit this calculation as supernova constraints on the cosmological parameters tighten toward the percent level.  The effect from a gravitational redshift would mimic a time-varying dark energy model even if we live in a universe where dark energy is a cosmological constant, and cause tensions between standard candle and standard ruler measurements. Thus it will be important for the next generation of SN Ia surveys to consider potential biases caused by gravitational redshifts or other redshift errors.

\begin{table}
\centering
\caption{Information criteria results with and without $\Delta z$ for fits to JLA alone and JLA$+$CMB.  These models are the ones shown in Fig.~\ref{fig:combinedres}.}
\label{tb:aicbic}
{\tabulinesep=0.8mm
\begin{tabu}{lccccccc}
\hline
\hline
 & $\chi^2$ & $k$ & $N$ & AIC & BIC & $\Delta$AIC & $\Delta$BIC \\ \hline
JLA & 682.7& 5 & 740 &   692.7 & 715.7 & 0 & 0 \\
JLA $\Delta z$ & 681.8 & 6 & 740 &  693.8 & 721.4 & -1.1 & -5.7 \\ \hline
JLA with CMB  & 683.0 & 5 & 741 &  693.0 & 716.0 & 0 & 0 \\
JLA $\Delta z$ with CMB & 681.8 & 6 & 741 & 693.8 & 721.4 & -0.8 & -5.4 \\ \hline
\end{tabu}}
\end{table}

\subsection{Interpretations of a redshift bias}
The value of $\Delta z$ we have found is much larger than expected for a gravitational redshift in the standard cosmology, and would amount to a Doppler shift caused by a velocity on the order of 100 \kms\ away from us. Even from the most extreme clusters, we would expect a gravitational redshift to be no more than $\Delta z \sim -10^{-4}$ \citep[see][Fig.~3]{wojtak2015}. If $\Delta z$ is positive it would mean that the observed redshift is higher than the cosmological one, and thus the light would have shifted more towards the red as it reached the observer than the purely cosmological redshift would allow. If this redshift bias was due to our gravitational environment, it would mean that we are located in an under-dense region in the cosmic web.

The magnitude of $\Delta z$ is more comparable with typical peculiar velocities, such as the heliocentric to CMB correction ($\sim370$\kms).  However, there are no mechanisms in the standard model to produce such velocities, unless our local under density is large enough to affect most of the low-$z$ supernovae in the sample.  This would be consistent with the Milky-way being in an under-density, as we expect outflow from under-densities as galaxies are gravitationally attracted to the more dense regions further away from us.  So both the gravitational redshift and the peculiar velocities would contribute extra positive redshift.  However, knowing the sizes of typical density fluctuations predicted in $\Lambda$CDM, we would expect that the supernovae in our sample are distant enough to be far from any local under density that could contribute a monopole term (a Hubble bubble).

If the redshift bias $\Delta z$ that we have fitted for in this paper is due to a local under-density, its size would have to be limited to the closest SN Ia samples in the JLA catalogue. We can only be sensitive to the biasing effects from gravitational redshifts if the supernovae are located outside of our local over or under-density. Since the low redshift supernovae provide most of the constraining power on $\Delta z$, it is fair to assume that they would be located in a different gravitational environment to our own, thus setting an upper limit on the size of our own gravitational environment.

For $z \ll 1$ we can find the distance to an object using
\begin{equation}
D = \frac{c}{100h}z.
\end{equation}
Choosing a redshift in the range of $z \sim 0.02-0.03$ to be the edge of the void would correspond to a distance of roughly 60 - 90 \hinvMpc. We can determine the density contrast between the void and the surrounding universe using the expression \cite{wojtak2015}
\begin{equation}
\delta_R = \frac{-2 z_{\phi} c^2}{\Omega_m 100^2h^2R^2},
\end{equation}
where $z_{\phi}$ is the gravitational redshift and $R$ is the radius of the void. For the best fit value of the redshift bias the expected density contrast between the void and the average density of the universe would be on the order of $\delta_R \sim  -4.3$ to $\delta_R\sim -2.0$. 

As a point of comparison, the profile of this void is more drastic than the super-void found in the CMB, where $R = (220 \pm 50)$ \hinvMpc \ and $\delta \approx -0.14 \pm 0.04$, which corresponds to a gravitational redshift of $z_{\phi} = 1.1\times 10^{-4}$ for an observer located at its centre \citep{szapudi2015}.

Note that this simple test of plausibility ignores the effects of peculiar velocities on the supernovae. Clearly if we were located in a large under-density, the peculiar velocities of host galaxies will also cause a large systematic bias. Nearby galaxies will be gravitationally attracted towards structure located outside of our local under-density, causing additional redshifting of their light. This signal would mix in with the gravitational effects from the local under-density, but should diminish as the low redshift cut off is increased.

Another possibility is that the $\Delta z$ signal is not due to extra-galactic effects, but something more mundane. It is possible to reproduce the appearance of a gravitational redshift or peculiar velocity bias by simply making systematic errors during observations. If redshifting software produces a systematic shift in the redshift values, say due to sub-pixel rounding errors, it could appear as a signal in the analysis stage. In fact, most of the heliocentric redshift measurements in the JLA sample are only quoted to 3 decimal places. If this is indeed the best accuracy with which the redshift could be determined, then when correcting to the CMB frame the inaccuracy would persist. The CMB redshift is used to define a range of integration in equation \eqref{eq:comoving}, and hence a small change in this value can significantly change the luminosity distance, particularly at low redshift. It is possible that the redshift bias that we have found is due solely to the lack of precision with which the heliocentric frame redshifts are measured.

Accurate redshifts are particularly important for low-redshift supernovae; $\Delta z$ gets most of its constraining power at low redshift, since the slope of the magnitude-redshift relation is steepest there and a small shift in redshift results in a large shift in magnitude \citep[see][Fig.~4]{wojtak2015}. Although the redshift bias is relatively small, its impact on the best fit matter density will inevitably affect the best fit values of parameters that correlate with the matter density. Such parameters as the growth rate of structure ($f$), the strength of clustering ($\sigma_8$), and Hubble's constant ($H_0$) are expected to deviate along with the matter density. Next we assess whether this could possibly explain the discrepancy between the value of Hubble's constant derived from SN Ia ($73.03 \pm 1.79$ kms$^{-1}$Mpc$^{-1}$ \citep{riess2016}) and the Planck 2015 result of $H_0 = 67.51 \pm 0.64$ kms$^{-1}$Mpc$^{-1}$ \cite{planck2015}.

\subsection{The dependence of Hubble's constant on a redshift bias}\label{sec:deponhub}

\begin{figure}
\centering
\includegraphics[width=1\textwidth]{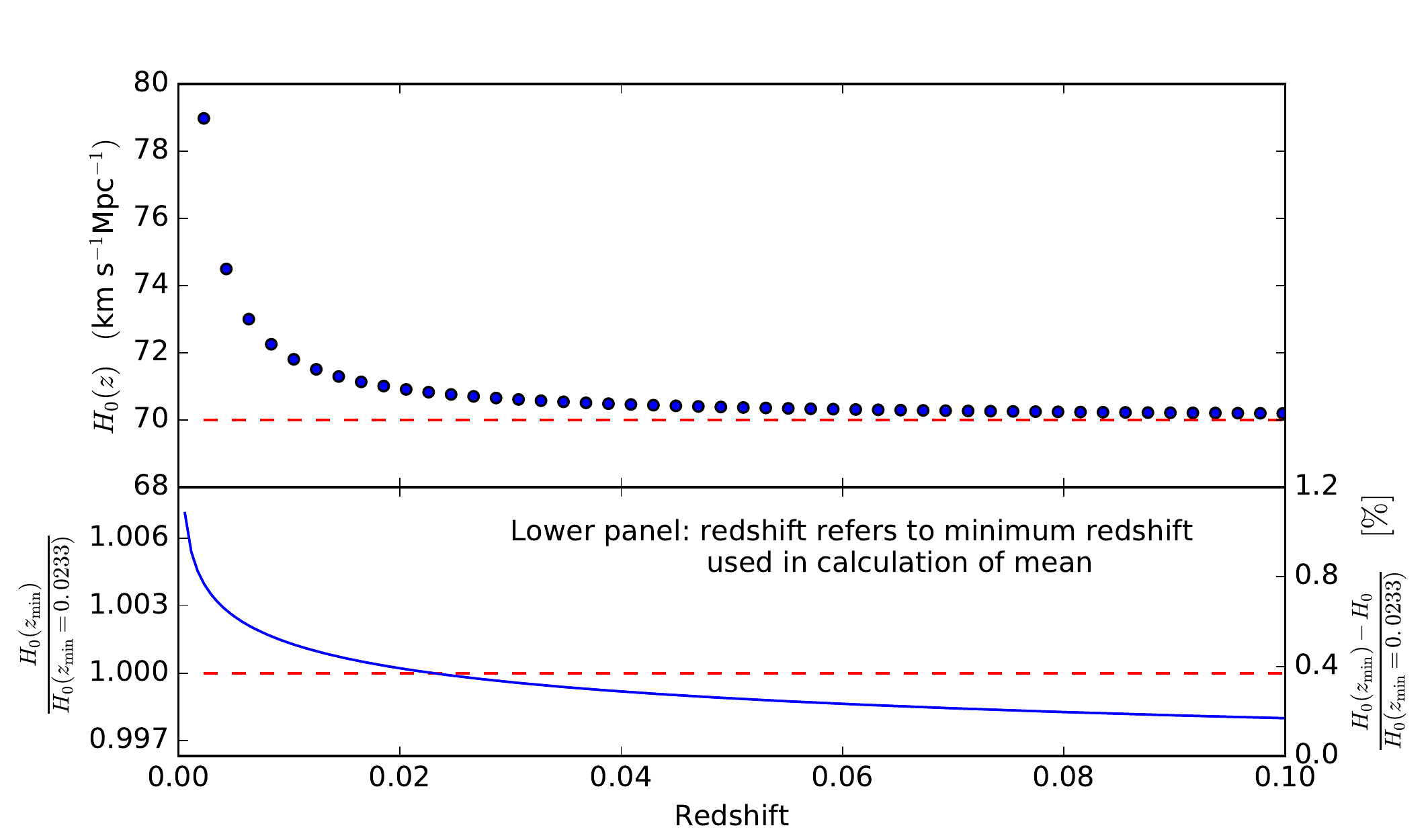}
\caption{ {\bf Upper:} Results of measurements of Hubble's constant from a sample of data that has a systematic bias in redshift of $\Delta z = 2.6\times10^{-4}$ and a fiducial value of $H_0 = 70$ kms$^{-1}$Mpc$^{-1}$.  The bias on $H_0$ is strongest for low-redshift objects.  This shows that a very small shift in redshift, can have a large impact on the derivation of the Hubble parameter.  {\bf Lower:}  Our prediction for direct comparison with Figure 13 of \cite{riess2016}.  They calculate $H_0$ by averaging over a window of 0.15 in redshift, where the minimum redshift is given on the horizontal axis.  We (and they) plot the ratio between $H_0$ as a function of the minimum redshift included in the calculation, and $H_0$ with minimum redshift set to $z_{\rm min}=0.0233$, i.e. $H_0(z_{\rm min})/H_0(z_{\rm min}\equiv 0.0233)$.   Since redshift bias has a greater impact on lower-redshift objects, the offset from the true value becomes smaller when low-redshift objects are excluded.  Our result shows that if redshift bias is responsible for a high value of $H_0$, then that should decrease as the minimum redshift increases, but \cite{riess2016} Fig.~13 shows the opposite trend.} 
\label{fig:H0_vs_z_inferred}
\end{figure}

Here we assess the impact a redshift bias would have on the measurement of the Hubble constant using nearby ($\bar{z}\lsim 0.1$) standard candles.  We create  a fake set of distance modulus data as a function of cosmological redshift, $\bar{z}$, for a flat $\Lambda$CDM model with $\Omega_m = 0.3$.  We then pretend that the redshifts we measure have been biased by $\Delta z$, so we infer $(1+z_{\rm cmb})= (1+\bar{z})(1+\Delta z)$. 
These redshift values are then used to determine Hubble's constant using
\begin{equation}
H_{0,i} = v(z_{\rm cmb, i})/D_i,
\end{equation}
where $D_i$ is the proper distance to the $i$th object, which is given by the measured luminosity distance according to $D_i = D_{{\rm L},i}/(1+z_{{\rm geo},i})$.  The velocity is not directly observable, so we infer it from the measured redshift $z_i$ according to,  
\beq v(z) = c z\left[1+\frac{1}{2}(1-q_0)z - \frac{1}{6}(1-q_0-3q_0^2+j_0)z^2\right], \label{eq:vzHD}\eeq 
following Riess et al.\ 2016 \cite{riess2016}, where the deceleration parameter $q_0=-0.55$ and jerk $j_0=1.0$ for standard $\Lambda$CDM with $(\om,\oll)\sim(0.3,0.7)$, \citep{riess2011,riess2016}.  This formula approximates velocity as a function of redshift well for $z\lsim 0.1$ for most viable cosmological models (it is is only very weakly sensitive to the cosmological parameters such as $\Omega_{\rm m}$ and $\Omega_{\Lambda}$).

The bias due to a redshift shift is strongest at low redshift, as seen in Fig.~\ref{fig:H0_vs_z_inferred}, where each dot represents an example measurement of $H_{0,i}$ from a single measurement of distance and redshift with $\Delta z=2.6\times 10^{-4}$.  Then the inferred value of $H_0$ for the sample is the average of all the $H_{0,i}$.  
We find that a sample evenly distributed in redshift out to $z<0.1$ would overestimate Hubble's constant by $\sim$1  kms$^{-1}$Mpc$^{-1}$ if there was a redshift bias of $\Delta z=2.6\times10^{-4}$.  Implementing a low-redshift cutoff to the data so that $0.0233<z<0.1$ would reduce the overestimation to $\sim$$0.3$ kms$^{-1}$Mpc$^{-1}$.

\subsection{Dependence on lowest redshift}

\begin{figure}
\centering
\includegraphics[width=0.7\textwidth]{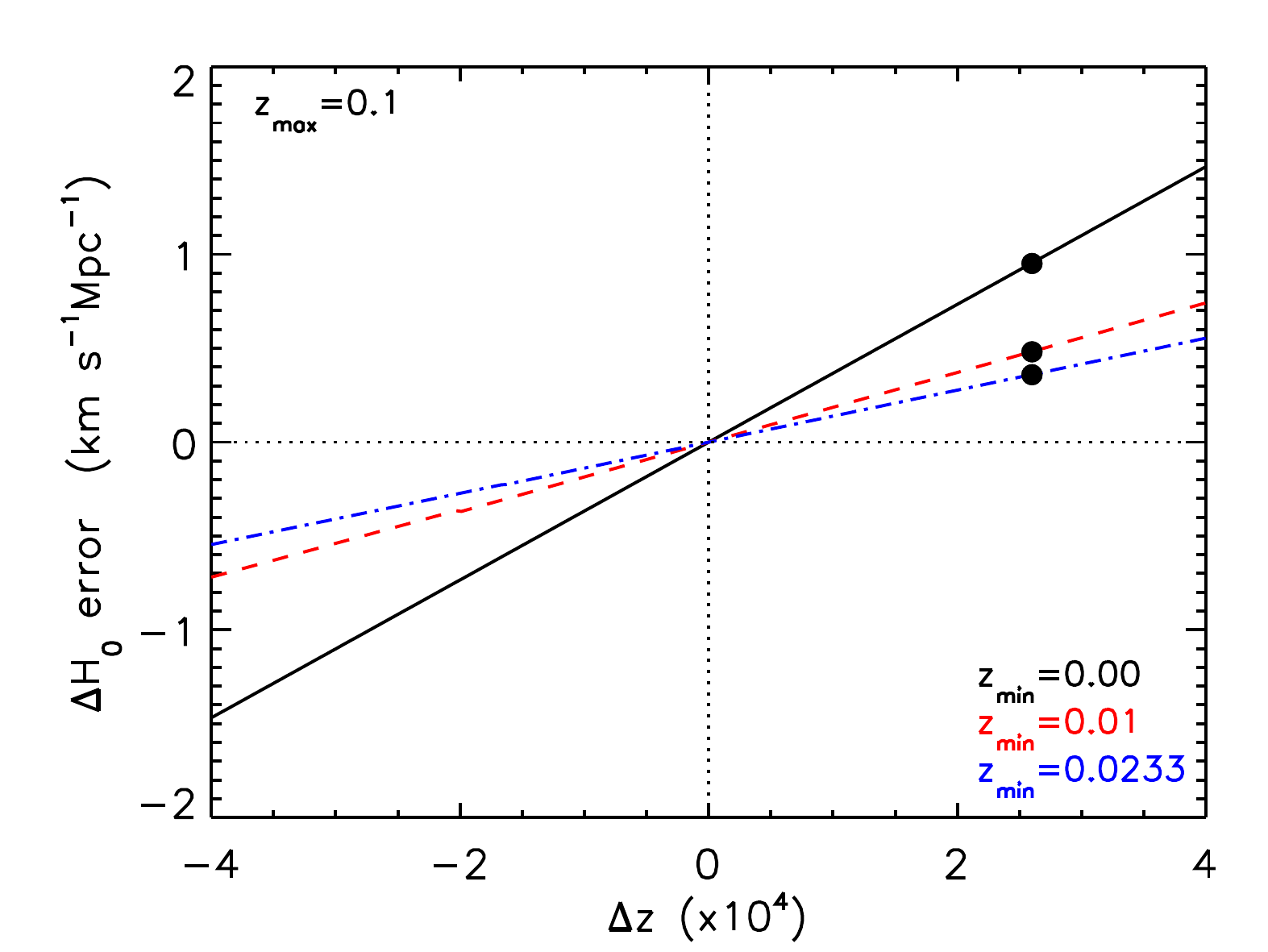}
\caption{Results of measurements of Hubble's constant from a sample of data that has a systematic bias in redshift given by the value on the horizontal axis.  Data points are spread evenly from $0.0<\zbar<0.1$. The observed redshift is given by the true cosmological redshift multiplied by a redshift bias as marked on the the horizontal axis, i.e. $z_{\rm obs}=(1+\zbar)(1+\Delta z)-1$.  Note the redshift axis has been multiplied by $10^{4}$, so the range is from $-4\times10^{-4} <z<4\times10^{-4}$.
The red dashed line and blue dot-dashed line show how the effect is diminished when you remove low-redshift data points, with $z_{\rm min}=0.01$ and $z_{\rm min}=0.0233$ respectively.   The solid point marks the error that would be made if $\Delta z = 2.6\times10^{-4}$ but it was assumed to be zero.}
\label{fig:H0zshift}
\end{figure}

Clearly the overall value of $H_0$ that is inferred will depend on the redshift range used in the average of $H_{0,i}$.  Riess et al.\ \cite{riess2016} test for redshift-range effects by using a sliding window of redshift width $0.15$ and then recalculate $H_0$ as they increase the lower end of that window from $z=0.0233$ to $z_{\rm min}=0.25$.  They find an increase in their inferred $H_0$ of approximately 0.05\% over that range.  

We repeat that test for our fake data with a redshift offset, and show the results in the lower panel of Figure~\ref{fig:H0_vs_z_inferred}.  We also find a shift of approximately 0.05\% with this test in the presence of a redshift bias of $\Delta z=2.6\times 10^{-4}$, but in the opposite direction to \cite{riess2016} (decreasing $H_0$ with increasing redshift window).  Therefore we conclude that although this effect is of the correct order of magnitude, its direction is in the opposite sense to that found by \cite{riess2016} and so is unlikely to be the cause of the discrepancy. Figure~\ref{fig:H0zshift} shows the offset in $H_0$ expected for a range of $\Delta z$ values and low-$z$ cutoffs.

Our results are in line with a more detailed analysis provided by \cite{odderskov2016}, who study the effect of inhomogeneities on the calculation of Hubble's constant using mock observations and $N$-body simulations. In their analysis they find that the sign of the change in Hubble's constant due to inhomogeneities is dependent on the observer being located in gravitationally over or under-dense regions in the cosmic web, where a higher local Hubble's constant is measured by observers in under-dense regions \cite{odderskov2016}.

\section{Conclusion}

In this paper we looked for evidence of a redshift bias in the Joint Light-curve Analysis SN Ia dataset by \cite{betoule2014}, either due to gravitational redshifts or systematic errors. We introduce the redshift bias parameter $\Delta z$, which is added as a free parameter to the fitting process. We find a best fit value of $\Delta z = (2.6^{+2.7}_{-2.8}) \times 10^{-4}$. If interpreted as a gravitational redshift, the best fit value indicates that we would be located in an under-dense region in the cosmic web.  Although this value is consistent with zero to within $1\sigma$, allowing this extra freedom in the fit changes the best fit value of matter density by just under $0.5\sigma$, from $\Omega_m = 0.296 ^{+0.037}_{-0.035}$ to $\Omega_m = 0.313^{+0.042}_{-0.040}$. The shifted matter density value closely matches the value obtained from the CMB in the Planck 2015 results, who find the value to be $\Omega_m = 0.3121 \pm 0.0087$ \cite{planck2015}. These shifts are not large enough to be a major concern for the current data set, but demonstrate the importance of accurate supernova redshifts when doing supernova cosmology.  

Our aim in doing this measurement was to test whether supernovae can be used to diagnose whether we live in a local under- or over-density.  Information criteria tests show that the current precision of the data is insufficient to support the inclusion of this extra parameter. However, it is clear that supernovae can be sensitive probes of the local environment when redshifts are measured sufficiently accurately. Current high redshift supernova samples typically have redshift uncertainties quoted to three decimal places, although roughly 90\% of the low redshift ($z<0.1$) supernovae have redshifts quoted with 4 to 5 decimal places. We have shown that systematic errors have to be controlled to the level of at least four decimal places (preferably five \cite{wojtak2015}) to avoid biasing cosmological parameters. This is mostly being done for the low redshift supernovae at present, however as supernovae numbers increase at higher redshifts the accuracy that they are quoted to will become more important if there exists some underlying systematic biases. 

The fact that the SN best-fit matter density aligns better with the measurement from the CMB after we allowed $\Delta z$ to vary could be interesting but is likely to be a statistical fluke.  If the $\Delta z$ is non-negligible and truly does originate due to an under-density then that would be consistent with local supernova measurements that indicate a higher Hubble constant than measured in the CMB. This is not due to local outflows, but because we lack knowledge of our local environment and do not account for it.

However, the best fit $\Delta z$ we measure exceeds the most extreme density fluctuation we would expect in $\Lambda$CDM.  While that does not preclude a larger-than-expected density fluctuation as the source (we are after all using these measurements to test $\Lambda$CDM), we conclude that a more likely explanation is a small systematic error in the data -- or simply a statistical fluctuation.  There are many ways in which a $10^{-4}$ systematic shift could have snuck into redshift measurements (software precision, rounding errors, pixel-to-wavelength mapping).
We therefore stress the importance of accurate redshift measurements when determining supernova cosmology. We recommend that the redshift values be quoted to at least 5 decimal values to reduce the possibility of contamination from systematic software or calculation biases.

\section{Acknowledgments}
We thank Krzysztof Bolejko for useful discussions. Parts of this research were conducted by the Australian Research Council Centre of Excellence for All-sky Astrophysics (CAASTRO), through project number CE110001020.

\newpage
\appendix
\section{Kinematic and cosmological redshifts in the definition of luminosity distance}

Luminosity distance ($D_L$) is a distance measure defined such that the observed flux ($F$: energy per area per unit time) obeys an inverse square law relationship with the luminosity ($L$: energy per unit time), 
\beq 
F = \frac{L}{4\pi D_L^2}.
\label{eq:lumdist}
\eeq
In flat, non-expanding space, luminosity distance would be equal to the proper distance.  In the expanding universe, which can also have curved space, the relationship is more complex.  
In the Friedmann-Robertson-Walker metric, defined by 
\beq
ds^2 = -c^2dt^2 + R(t)^2 \left[d\chi^2 + S_k^2(\chi)\left(d\theta^2 + \sin^2\theta d\phi^2\right)\right] ,
\eeq
where $S_k(\chi)=\sin\chi, \chi, \sinh\chi$ in closed, flat and open universes respectively, the infinitesimal surface area elements (along a surface with $dt=d\chi=0$) are $ds_\theta = R(t)S_k(\chi)d\theta$ and $ds_\phi=R(t)S_k(\chi)\sin\theta d\phi$.  Combining these to get the surface area gives,
\bea
A(t,\chi(z)) &=& R^2(t)S_k^2(\chi)\int_0^{2\pi}d\phi \int_{-\pi/2}^{\pi/2} \sin\theta d\theta , \\
 	&=& 4\pi R^2(t)S_k^2(\chi(z)).
\eea
The time of relevance is the time of observation, so $R(t)=R_0$.  Moreover, this area is entirely dependent on how much the universe has expanded between the time of emission and observation, and doesn't care about how fast the emitter and observer were travelling, nor their relative environmental densities.\footnote{ In detail there will be beaming and focussing of the light, and so the effective area at the point of observation does care about these things.  However, in the way we've set up the problem with all inhomogeneities being small perturbations on the homogeneous background, all those extra effects are taking into account by peculiar velocities and gravitational redshifts.}   So the redshift that should go into calculating the comoving coordinate is the cosmological redshift, $\bar{z}$.  Thus,
\beq 
A(\bar{z}) = 4\pi R_0^2 S_k^2(\chi(\bar{z})) 
\eeq 
The flux is also diminished by two other factors.  Firstly, the light is redshifted, and therefore the energy diminished by a factor of $(1+z)$.  Secondly, there is time dilation, which contributes another factor of $(1+z)$.  Both the energy loss and the time dilation are related to the actual redshift the light experiences, no matter how that redshift was generated.  So peculiar velocities and gravitational potentials, which change the redshift, also change both these factors of $(1+z)$.  Therefore the redshift in these terms needs to be $z_{\rm obs}$.

As a result of all these effects, the flux that is actually observed is
\beq F=\frac{L}{4\pi R_0^2 S_k^2(\chi(\bar{z})) (1+z_{\rm obs})^2}.\eeq
Equating this with Eq.~\ref{eq:lumdist} gives
\beq 
D_L(\bar{z},z_{\rm obs}) = R_0 S_k(\chi(\bar{z})) (1+z_{\rm obs}).
\eeq

\begin{figure}
	\begin{center}
	\includegraphics[width=0.7\textwidth]{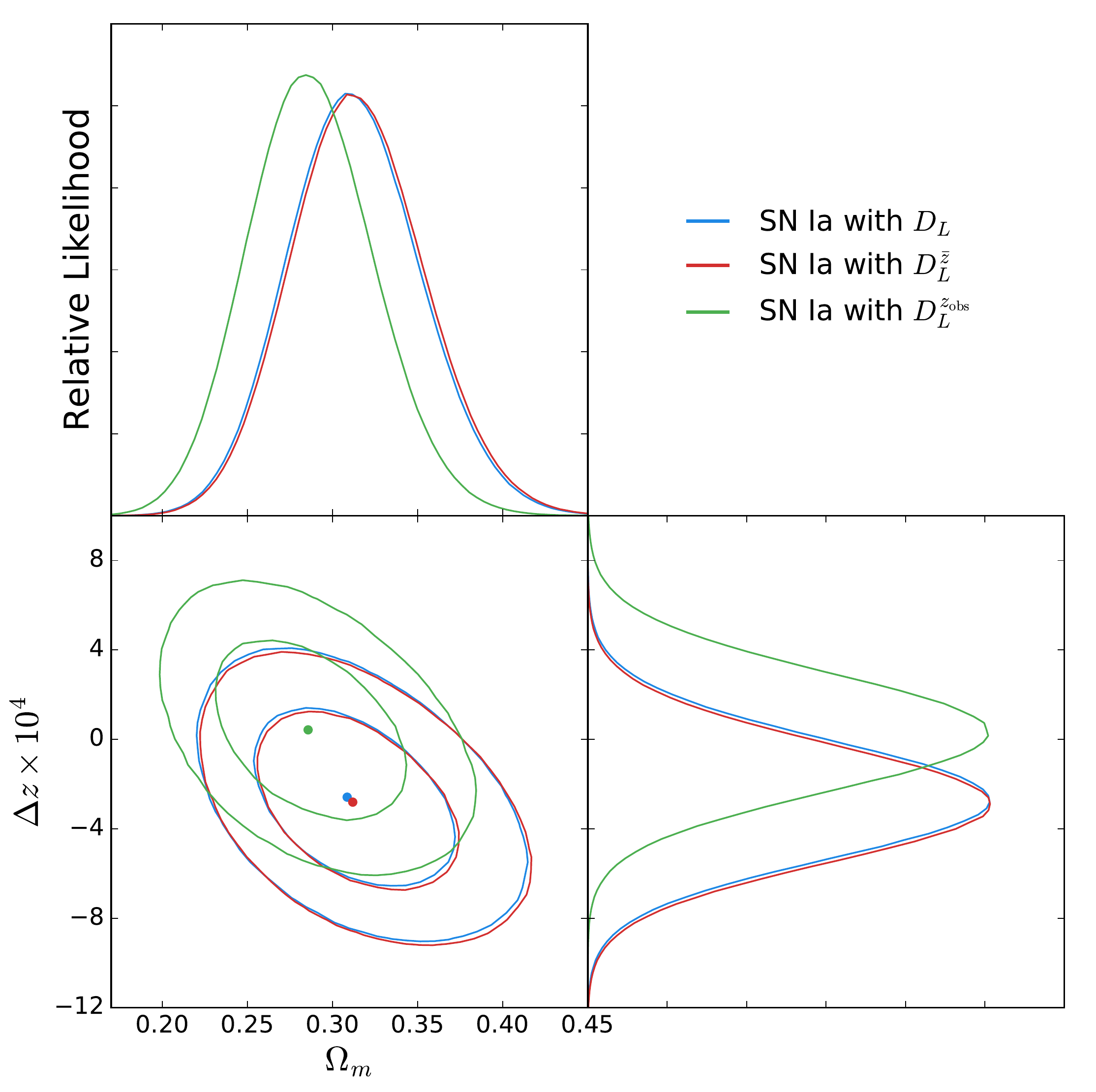}
	\end{center}
	\caption{Plot demonstrating the importance of using the correct redshifts (blue) in the equation for luminosity distance (Eq.~\ref{eq:DLcorrect}). Using only heliocentric redshifts (green) gives a clear and significant error -- which is well known and is why redshifts are (almost) always converted to the CMB frame before doing cosmological fits.  Using only CMB redshifts (red) is the most common way to calculate luminosity distance, and even with the JLA sample only causes a marginal difference compared to the correct method, which uses a combination of heliocentric and CMB redshifts.  With future samples of thousands of supernovae it may become important to take into account the difference between $z_{\rm geo}$ and $\bar{z}$ in the equation for luminosity distance to avoid bias in cosmological inferences.  }
	\label{fig:dL}
\end{figure}

In Figure~\ref{fig:dL} we show the impact on the $\Lambda$CDM cosmological fit when luminosity distance is calculated using various redshifts.  Not correcting redshifts to the CMB frame ($\bar{z}$) has a well-known important effect.  Until now it has been safe to use $\bar{z}$ in all terms in $D_L$, and indeed the difference will only become important if the precision of the measurements is improved by almost an order of magnitude.  Nevertheless, with supernova surveys soon to deliver thousands of supernovae we should pay attention to these details.

\providecommand{\href}[2]{#2}\begingroup\raggedright\endgroup


\begin{thebibliography}{10}

\bibitem{scrimgeour12}
M.~I. {Scrimgeour}, T.~{Davis}, C.~{Blake}, J.~B. {James}, G.~B. {Poole},
  L.~{Staveley-Smith}, S.~{Brough}, M.~{Colless}, C.~{Contreras}, W.~{Couch} et al.,
  {\it {The
  WiggleZ Dark Energy Survey: the transition to large-scale cosmic
  homogeneity}},  {\em \mnras} {\bf 425} (Sept., 2012) 116--134,
  [\href{http://arxiv.org/abs/1205.6812}{{\tt arXiv:1205.6812}}].

\bibitem{lavallaz2011}
A.~{de Lavallaz} and M.~{Fairbairn}, {\it {Effects of voids on the
  reconstruction of the equation of state of dark energy}},  {\em \prd} {\bf
  84} (Oct., 2011) 083005, [\href{http://arxiv.org/abs/1106.1611}{{\tt
  arXiv:1106.1611}}].

\bibitem{marra2013}
V.~{Marra}, M.~{P{\"a}{\"a}kk{\"o}nen}, and W.~{Valkenburg}, {\it {Uncertainty
  on w from large-scale structure}},  {\em \mnras} {\bf 431} (May, 2013)
  1891--1902, [\href{http://arxiv.org/abs/1203.2180}{{\tt arXiv:1203.2180}}].

\bibitem{yu2013}
B.~{Yu}, {\it {Impact of the local void on the cosmological parameters}},  {\em
  \jcap} {\bf 3} (Mar., 2013) 013.

\bibitem{valkenburg2013}
W.~{Valkenburg}, M.~{Kunz}, and V.~{Marra}, {\it {Intrinsic uncertainty on the
  nature of dark energy}},  {\em Physics of the Dark Universe} {\bf 2} (Dec.,
  2013) 219--223, [\href{http://arxiv.org/abs/1302.6588}{{\tt
  arXiv:1302.6588}}].

\bibitem{wojtak2015}
R.~{Wojtak}, T.~M. {Davis}, and J.~{Wiis}, {\it {Local gravitational redshifts
  can bias cosmological measurements}},  {\em \jcap} {\bf 7} (July, 2015) 25,
  [\href{http://arxiv.org/abs/1504.00718}{{\tt arXiv:1504.00718}}].

\bibitem{wojtak2011}
R.~{Wojtak}, S.~H. {Hansen}, and J.~{Hjorth}, {\it {Gravitational redshift of
  galaxies in clusters as predicted by general relativity}},  {\em \nat} {\bf
  477} (Sept., 2011) 567--569, [\href{http://arxiv.org/abs/1109.6571}{{\tt
  arXiv:1109.6571}}].

\bibitem{dele2012}
M.~J. {de Le{\'o}n Dom{\'{\i}}nguez Romero}, D.~{Garc{\'{\i}}a Lambas}, and
  H.~{Muriel}, {\it {An improved method for the identification of galaxy
  systems: Measuring the gravitational redshift by Dark Matter Haloes}},  {\em
  ArXiv e-prints} (Aug., 2012) [\href{http://arxiv.org/abs/1208.3471}{{\tt
  arXiv:1208.3471}}].

\bibitem{sadeh2015}
I.~{Sadeh}, L.~L. {Feng}, and O.~{Lahav}, {\it {Gravitational Redshift of
  Galaxies in Clusters from the Sloan Digital Sky Survey and the Baryon
  Oscillation Spectroscopic Survey}},  {\em Physical Review Letters} {\bf 114}
  (Feb., 2015) 071103, [\href{http://arxiv.org/abs/1410.5262}{{\tt
  arXiv:1410.5262}}].

\bibitem{boss2016}
S.~{Alam}, M.~{Ata}, S.~{Bailey}, F.~{Beutler}, D.~{Bizyaev}, J.~A. {Blazek},
  A.~S. {Bolton}, J.~R. {Brownstein}, A.~{Burden}, C.-H. {Chuang} et al., {\it
  {The clustering of galaxies in the completed SDSS-III Baryon Oscillation
  Spectroscopic Survey: cosmological analysis of the DR12 galaxy sample}},
  {\em ArXiv e-prints} (July, 2016)
  [\href{http://arxiv.org/abs/1607.03155}{{\tt arXiv:1607.03155}}].

\bibitem{betoule2014}
M.~{Betoule}, R.~{Kessler}, J.~{Guy}, J.~{Mosher}, D.~{Hardin}, R.~{Biswas},
  P.~{Astier}, P.~{El-Hage}, M.~{Konig}, S.~{Kuhlmann} et al., {\it {Improved
  cosmological constraints from a joint analysis of the SDSS-II and SNLS
  supernova samples}},  {\em \aap} {\bf 568} (Aug., 2014) A22,
  [\href{http://arxiv.org/abs/1401.4064}{{\tt arXiv:1401.4064}}].

\bibitem{planck2015}
{Planck Collaboration}, P.~A.~R. {Ade}, N.~{Aghanim}, M.~{Arnaud},
  M.~{Ashdown}, J.~{Aumont}, C.~{Baccigalupi}, A.~J. {Banday}, R.~B.
  {Barreiro}, J.~G. {Bartlett}, and et~al., {\it {Planck 2015 results. XIII.
  Cosmological parameters}},  {\em ArXiv e-prints} (Feb., 2015)
  [\href{http://arxiv.org/abs/1502.01589}{{\tt arXiv:1502.01589}}].

\bibitem{riess2016}
A.~G. {Riess}, L.~M. {Macri}, S.~L. {Hoffmann}, D.~{Scolnic}, S.~{Casertano},
  A.~V. {Filippenko}, B.~E. {Tucker}, M.~J. {Reid}, D.~O. {Jones}, J.~M.
  {Silverman} et al., {\it
  {A 2.4\% Determination of the Local Value of the Hubble Constant}},  {\em
  ArXiv e-prints} (Apr., 2016) [\href{http://arxiv.org/abs/1604.01424}{{\tt
  arXiv:1604.01424}}].

\bibitem{calcino2015}
J.~{Calcino}, ``{The Effect of Redshift Bias in Cosmology}.'' unpublished
  thesis, 2015.

\bibitem{sako2014}
M.~{Sako}, B.~{Bassett}, A.~C. {Becker}, P.~J. {Brown}, H.~{Campbell},
  R.~{Cane}, D.~{Cinabro}, C.~B. {D'Andrea}, K.~S. {Dawson}, F.~{DeJongh} et al., {\it
  {The Data Release of the Sloan Digital Sky Survey-II Supernova Survey}},
  {\em ArXiv e-prints} (Jan., 2014) [\href{http://arxiv.org/abs/1401.3317}{{\tt
  arXiv:1401.3317}}].

\bibitem{astier2006}
P.~{Astier}, J.~{Guy}, N.~{Regnault}, R.~{Pain}, E.~{Aubourg}, D.~{Balam},
  S.~{Basa}, R.~G. {Carlberg}, S.~{Fabbro}, D.~{Fouchez} et al.,
   {\it {The Supernova Legacy Survey: measurement of
  {$\Omega$}$_{M}$, {$\Omega$}$_{Λ}$ and w from the first year data set}},
  {\em \aap} {\bf 447} (Feb., 2006) 31--48,
  [\href{http://arxiv.org/abs/astro-ph/0510447}{{\tt astro-ph/0510447}}].

\bibitem{sullivan2011}
M.~{Sullivan}, J.~{Guy}, A.~{Conley}, N.~{Regnault}, P.~{Astier}, C.~{Balland},
  S.~{Basa}, R.~G. {Carlberg}, D.~{Fouchez}, D.~{Hardin} et. al, {\it {SNLS3: Constraints on
  Dark Energy Combining the Supernova Legacy Survey Three-year Data with Other
  Probes}},  {\em \apj} {\bf 737} (Aug., 2011) 102,
  [\href{http://arxiv.org/abs/1104.1444}{{\tt arXiv:1104.1444}}].

\bibitem{riess2007}
A.~G. {Riess}, L.-G. {Strolger}, S.~{Casertano}, H.~C. {Ferguson},
  B.~{Mobasher}, B.~{Gold}, P.~J. {Challis}, A.~V. {Filippenko}, S.~{Jha},
  W.~{Li} et al., {\it {New Hubble Space Telescope Discoveries
  of Type Ia Supernovae at z > 1: Narrowing Constraints on the Early Behavior
  of Dark Energy}},  {\em \apj} {\bf 659} (Apr., 2007) 98--121,
  [\href{http://arxiv.org/abs/astro-ph/0611572}{{\tt astro-ph/0611572}}].

\bibitem{suzuki2012}
N.~{Suzuki}, D.~{Rubin}, C.~{Lidman}, G.~{Aldering}, R.~{Amanullah},
  K.~{Barbary}, L.~F. {Barrientos}, J.~{Botyanszki}, M.~{Brodwin},
  N.~{Connolly} et al., {\it {The
  Hubble Space Telescope Cluster Supernova Survey. V. Improving the Dark-energy
  Constraints above z > 1 and Building an Early-type-hosted Supernova Sample}},
   {\em \apj} {\bf 746} (Feb., 2012) 85,
  [\href{http://arxiv.org/abs/1105.3470}{{\tt arXiv:1105.3470}}].

\bibitem{hicken2009}
M.~{Hicken}, W.~M. {Wood-Vasey}, S.~{Blondin}, P.~{Challis}, S.~{Jha}, P.~L.
  {Kelly}, A.~{Rest}, and R.~P. {Kirshner}, {\it {Improved Dark Energy
  Constraints from \~{}100 New CfA Supernova Type Ia Light Curves}},  {\em
  \apj} {\bf 700} (Aug., 2009) 1097--1140,
  [\href{http://arxiv.org/abs/0901.4804}{{\tt arXiv:0901.4804}}].

\bibitem{contreras2010}
C.~{Contreras}, M.~{Hamuy}, M.~M. {Phillips}, G.~{Folatelli}, N.~B. {Suntzeff},
  S.~E. {Persson}, M.~{Stritzinger}, L.~{Boldt}, S.~{Gonz{\'a}lez},
  W.~{Krzeminski} et al., {\it {The Carnegie Supernova
  Project: First Photometry Data Release of Low-Redshift Type Ia Supernovae}},
  {\em \aj} {\bf 139} (Feb., 2010) 519--539,
  [\href{http://arxiv.org/abs/0910.3330}{{\tt arXiv:0910.3330}}].

\bibitem{folatelli2010}
G.~{Folatelli}, M.~M. {Phillips}, C.~R. {Burns}, C.~{Contreras}, M.~{Hamuy},
  W.~L. {Freedman}, S.~E. {Persson}, M.~{Stritzinger}, N.~B. {Suntzeff},
  K.~{Krisciunas} et al., {\it {The Carnegie
  Supernova Project: Analysis of the First Sample of Low-Redshift Type-Ia
  Supernovae}},  {\em \aj} {\bf 139} (Jan., 2010) 120--144,
  [\href{http://arxiv.org/abs/0910.3317}{{\tt arXiv:0910.3317}}].

\bibitem{stritzinger2011}
M.~D. Stritzinger, M.~M. Phillips, L.~N. Boldt, C.~Burns, A.~Campillay,
  C.~Contreras, S.~Gonzalez, G.~Folatelli, N.~Morrell, W.~Krzeminski et al., 
  {\it The carnegie supernova project: Second photometry
  data release of low-redshift type ia supernovae},  {\em The Astronomical
  Journal} {\bf 142} (2011), no.~5 156.

\bibitem{ganeshalingam2013}
M.~{Ganeshalingam}, W.~{Li}, and A.~V. {Filippenko}, {\it {Constraints on dark
  energy with the LOSS SN Ia sample}},  {\em \mnras} {\bf 433} (Aug., 2013)
  2240--2258, [\href{http://arxiv.org/abs/1307.0824}{{\tt arXiv:1307.0824}}].

\bibitem{aldering2002}
G.~{Aldering}, G.~{Adam}, P.~{Antilogus}, P.~{Astier}, R.~{Bacon},
  S.~{Bongard}, C.~{Bonnaud}, Y.~{Copin}, D.~{Hardin}, F.~{Henault} et al., {\it {Overview of the
  Nearby Supernova Factory}},  in {\em Survey and Other Telescope Technologies
  and Discoveries} (J.~A. {Tyson} and S.~{Wolff}, eds.), vol.~4836 of {\em
  \procspie}, pp.~61--72, Dec., 2002.

\bibitem{phillips1993}
M.~M. {Phillips}, {\it {The absolute magnitudes of Type IA supernovae}},  {\em
  \apjl} {\bf 413} (Aug., 1993) L105--L108.

\bibitem{phillips1999}
M.~M. {Phillips}, P.~{Lira}, N.~B. {Suntzeff}, R.~A. {Schommer}, M.~{Hamuy},
  and J.~{Maza}, {\it {The Reddening-Free Decline Rate Versus Luminosity
  Relationship for Type IA Supernovae}},  {\em \aj} {\bf 118} (Oct., 1999)
  1766--1776, [\href{http://arxiv.org/abs/astro-ph/9907052}{{\tt
  astro-ph/9907052}}].

\bibitem{tripp1998}
R.~{Tripp}, {\it {A two-parameter luminosity correction for Type IA
  supernovae}},  {\em \aap} {\bf 331} (Mar., 1998) 815--820.

\bibitem{schwarz1978}
G.~Schwarz, {\it Estimating the dimension of a model},  {\em Ann. Statist.}
  {\bf 6} (03, 1978) 461--464.

\bibitem{akaike1974}
H.~Akaike, {\it A new look at the statistical model identification},  {\em IEEE
  Transactions on Automatic Control} {\bf 19} (Dec, 1974) 716--723.

\bibitem{szapudi2015}
I.~{Szapudi}, A.~{Kov{\'a}cs}, B.~R. {Granett}, Z.~{Frei}, J.~{Silk},
  W.~{Burgett}, S.~{Cole}, P.~W. {Draper}, D.~J. {Farrow}, N.~{Kaiser} et al., 
  {\it {Detection of a supervoid aligned with the cold spot of
  the cosmic microwave background}},  {\em \mnras} {\bf 450} (June, 2015)
  288--294, [\href{http://arxiv.org/abs/1405.1566}{{\tt arXiv:1405.1566}}].

\bibitem{riess2011}
A.~G. {Riess}, L.~{Macri}, S.~{Casertano}, H.~{Lampeitl}, H.~C. {Ferguson},
  A.~V. {Filippenko}, S.~W. {Jha}, W.~{Li}, and R.~{Chornock}, {\it {A 3\%
  Solution: Determination of the Hubble Constant with the Hubble Space
  Telescope and Wide Field Camera 3}},  {\em \apj} {\bf 730} (Apr., 2011) 119,
  [\href{http://arxiv.org/abs/1103.2976}{{\tt arXiv:1103.2976}}].

\bibitem{odderskov2016}
I.~{Odderskov}, S.~M. {Koksbang}, and S.~{Hannestad}, {\it {The local value of
  H$_{0}$ in an inhomogeneous universe}},  {\em \jcap} {\bf 2} (Feb., 2016)
  001, [\href{http://arxiv.org/abs/1601.07356}{{\tt arXiv:1601.07356}}].

\end{thebibliography}
\end{document}